\documentclass[aps,prstab,reprint,groupedaddress,showpacs,amsmath,amssymb]{revtex4-1}
\usepackage{epsfig}
\usepackage{dcolumn}
\usepackage{bm}

\begin{document}


\title{Decoupling of beams previously coupled by effective stand-alone solenoid fringe fields}


\author{L.~Groening}
\affiliation{GSI Helmholtzzentrum f\"ur Schwerionenforschung GmbH, Darmstadt D64291, Germany}


\date{\today}

\begin{abstract}
Beams passing through a solenoid fringe field experience x-y coupling and change of their eigen-emittances. As reported previously (C.~Xiao et al., Phys. Rev. ST Accel. Beams 044201, {\bf 16} 2013) constant settings of a subsequent decoupling section can be found such that variation of the fringe field strength will not change the Twiss parameters $\beta$ and $\alpha$ in both transverse planes at the exit of the decoupling section. For time being this feature was understood for a generic beam line but not to the generality to which it is observed. This report is on explanation of the convenient decoupling of fringe-coupled beams by any beam line that provides decoupling. For better coherence this report includes recapitulation of previous works.
\end{abstract}

\pacs{41.75.Ak, 41.85.Ct, 41.85.Ja}

\maketitle


\section{Introduction}
Transformation of a round beam (equal transverse emittances) to a flat beam (different transverse emittances) requires changing the beam eigen-emittances. The eigen-emittances are defined through the beam second moments as
\begin{equation}
{\varepsilon_1}=\frac{1}{2} \sqrt{-tr[(CJ)^2] + \sqrt{tr^2[(CJ)^2]-16 det(C) }}
\end{equation}
\begin{equation}
{\varepsilon_2}=\frac{1}{2} \sqrt{-tr[(CJ)^2] - \sqrt{tr^2[(CJ)^2]-16 det(C) }},
\end{equation}
where
\begin{equation}
\label{2nd_mom_matrix}
C=
\begin{bmatrix}
\langle xx \rangle &  \langle xx'\rangle &  \langle xy\rangle & \langle xy'\rangle \\
\langle x'x\rangle &  \langle x'x'\rangle & \langle x'y\rangle & \langle x'y'\rangle \\
\langle yx\rangle &  \langle yx'\rangle &  \langle yy\rangle & \langle yy'\rangle \\
\langle y'x\rangle &  \langle y'x'\rangle & \langle y'y\rangle & \langle y'y'\rangle
\end{bmatrix}
\end{equation}
and
\begin{equation}
J=
\begin{bmatrix}
0 &  1 &  0 & 0 \\
-1 &  0 &  0 & 0 \\
0 &  0 &  0 & 1 \\
0 &  0 & -1 & 0
\end{bmatrix}.
\end{equation}
Linear transport elements as drifts, quadrupoles, dipoles, and rf-gaps do not change neither the beam rms emittances nor the eigen-emittances. Solenoids, skew quadrupoles, and -dipoles change the rms emittances through x-y coupling. But they do not change the eigen-emittances. This is often expressed by the simplecticity criterion for the transport matrix $M$ representing the transport element~\cite{Dragt}
\begin{equation}
\label{symplecticity}
M^TJM=J.
\end{equation}
A matrix $M$ satisfying the above criterion, is called symplectic and the eigen-emittances of a beam being transported by $M$ remain constant. Beam particle coordinates are expressed by displacements x and y in space and by the respective derivatives x' and y' w.r.t. the longitudinal coordinate s.\\
The matrix of a solenoid fringe field reads as
\begin{equation}
\label{fringe}
M_{F}=
\begin{bmatrix}
1 &  0 &  0 & 0 \\
0 &  1 &  k & 0 \\
0 &  0 &  1 & 0 \\
-k &  0 & 0 & 1
\end{bmatrix}
\end{equation}
with $k=\frac{B}{2(B \rho)}$. $B$ is the solenoid on-axis magnetic field strength and $B\rho$ is the beam rigidity. $M_{F}$ does not satisfy Equ.~\ref{symplecticity} and changes the eigen-emittances. However, it leaves constant the 4d rms emittance defined as the square root of the determinant of $C$ from Equ.~\ref{2nd_mom_matrix}.
\\
Stand-alone fringe fields do not exist since magnetic field lines are closed. Effective stand-alone fringe fields act on the beam if the beam charge state is changed in between the fringes of the same solenoid. This is the case for rf-guns~\cite{Brinkmann,Edwards} (free electron creation inside solenoid), extraction from an Electron-Cyclotron-Resonance ion source~\cite{Bertrand} (ionisation inside the solenoid), and for charge state stripping inside a solenoid~\cite{Groening}. Further discussion of symplecticity of fringes shall be avoided here and we refer to~\cite{Baumgarten} instead. We just point out that changing the ion beam charge state is equivalent to cancelling the stripped-off electrons from the system. This cancellation is a non-symplectic action and conservation of the eigen-emittances within the remaining subsystem cannot by assumed in general.
\\
In this report we assume that an effective fringe field (Equ.~\ref{fringe}) coupled an initially round \& decoupled beam. The second moments matrix of this beam at the entrance to that fringe is given by
\begin{equation}
C_1^{'}=
\begin{bmatrix}
\varepsilon \beta &  0 &  0 & 0 \\
0 &  \frac{\varepsilon} {\beta} &  0 & 0 \\
0 &  0 &  \varepsilon \beta & 0 \\
0 &  0 & 0 & \frac{\varepsilon} {\beta}
\end{bmatrix}\,,
\end{equation}
where $\varepsilon$ is the rms emittance in both transverse planes and $\beta$ is the rms beta function.\\
The report is organized in the following: in the first section we repeat parts of references \cite{Kim} and \cite{Xiao_prstab}, i.e. decoupling of the beam using a generic decoupling beam line. The decoupling capabilities are derived for this case. We recapitulate the findings of \cite{Xiao_prstab} that any decoupling beam line seems to inhabit very convenient decoupling features. The subsequent section treats the extension of the generic case to any decoupling beam line, i.e. any decoupling beam line performs with the same convenient decoupling features as the generic beam line.

\section{De-coupling for the generic case}
\begin{figure}[hbt]
\centering
\includegraphics*[width=80mm,clip=]{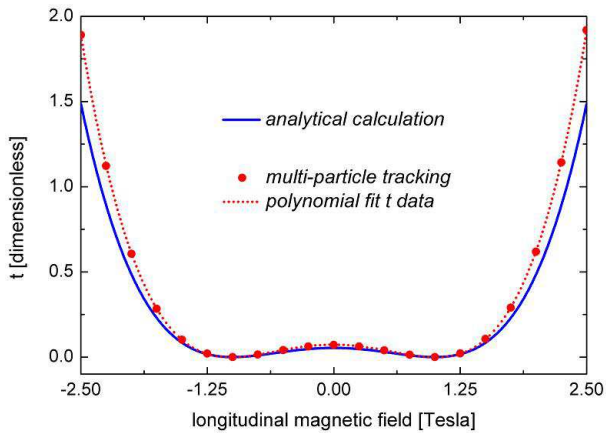}
\caption{The coupling parameter $t$ at the exit of the generic beam line $\overline{R}$ as a function of the solenoid field causing the fringe field strength $k_1$ (blue line). The figure is taken from~\cite{Xiao_prstab} and $k_0$ corresponds to 1.0~T. The dependency is described by Equ.~\ref{t value}.}
\label{t-graph}
\end{figure}

\begin{figure}[hbt]
\centering
\includegraphics*[width=80mm,clip=]{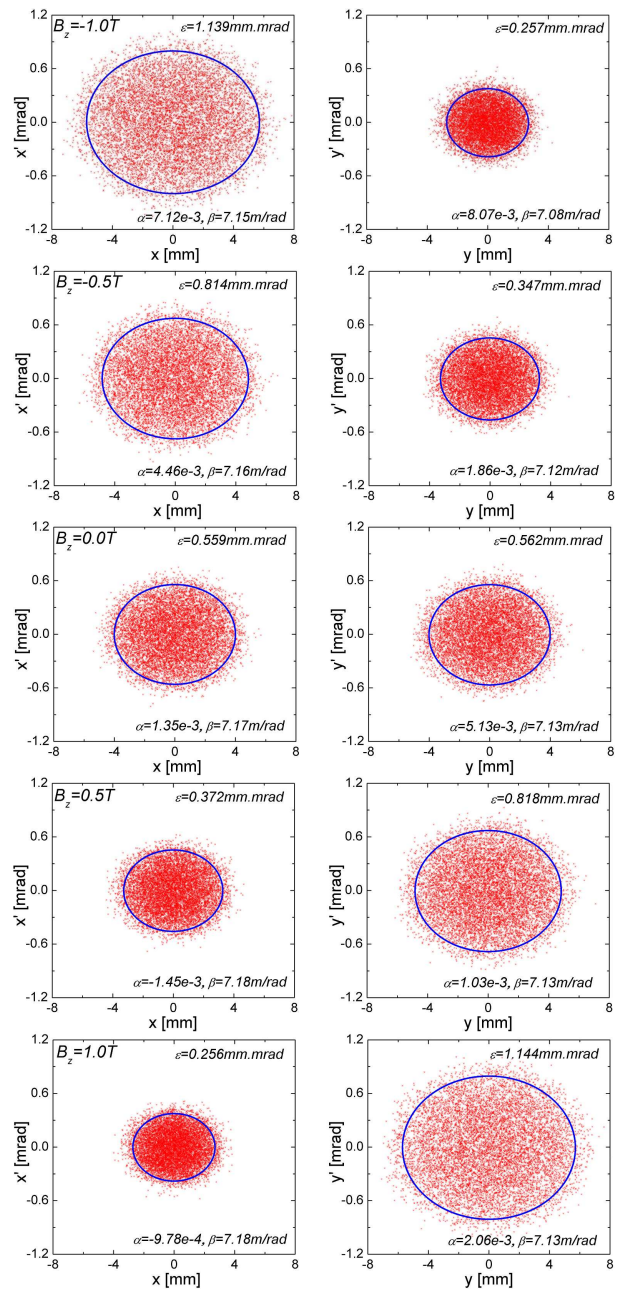}
\caption{Phase space distributions of beams initially coupled by different solenoid fringe fields and decoupled by the same gradients along the subsequent decoupling line. The figure is taken from~\cite{Xiao_prstab} and $k_0$ corresponds to 1.0~T. The ellipse parameters $\beta$ and $\alpha$ do not depend on the fringe field strength as suggested by Equ.~\ref{Twiss_const}.}
\label{ellipses}
\end{figure}

The beam second moment matrix after passing the fringe field of Equ.~\ref{fringe} is
\begin{equation}
\label{coupled}
{C_{2}^{'}=M_{F} C_{1}^{'} M_{F}^{T}=}
\begin{bmatrix}
\varepsilon_n R_n & -k\varepsilon_n \beta_n J_n \\
k\varepsilon_n \beta_n J_n & \varepsilon_n R_n
\end{bmatrix}\,,
\end{equation}
where
\begin{equation}
\varepsilon_n=\sqrt{ \varepsilon \beta (\frac{\varepsilon}{\beta}+k^2 \varepsilon \beta)}, ~~~\beta_n=\frac{\beta \varepsilon}{\varepsilon_n}\,,
\end{equation}
introducing the 2$\times$2 sub-matrices $R_n$ and $J_n$ as
\begin{equation}
R_n=
\begin{bmatrix}
\beta_n &  0  \\
0 &  \frac{1}{\beta_n}
\end{bmatrix}\,
,~~~
J_n=
\begin{bmatrix}
0 &  1  \\
-1 &  0
\end{bmatrix}.
\end{equation}
Inter-plane coupling is created and the rms emittances and eigen-emittances after the fringe read
\begin{equation}
\varepsilon_{x,y}=\varepsilon_n, ~~~\varepsilon_{1,2}=\varepsilon_n(1\mp k\beta_n)\,.
\end{equation}
\\
The parameter $t$ is introduced to quantify the interplane coupling. If $t$
defined as
\begin{equation}
\label{eq4}
t=\frac{\varepsilon_x \varepsilon_y}{\varepsilon_1 \varepsilon_2}-1\,\geq\,0
\end{equation}
is equal to zero, there are no inter-plane correlations and the beam is fully
decoupled.\\
Obtaining this result we neglected the finite solenoid length, i.e. its central longitudinal field. Tracking simulations using 3D-field maps of finite solenoids confirmed that this omission is justified~\cite{Xiao_HB2012}.\\
As shown for instance by Kim~\cite{Kim} the beam represented by Equ.~\ref{coupled} is decoupled through a beam line formed by an identity matrix in the $x$-direction and an additional 90$^\circ$ phase advance in $y$-direction
\begin{equation}
R_q=
\begin{bmatrix}
I_n &  O_n  \\
O_n &  T_n
\end{bmatrix}\,.
\end{equation}
Here the 2$\times$2 sub-matrices $O_n$, $T_n$ and $I_n$ are defined as
\begin{equation}
O_n=
\begin{bmatrix}
0 &  0  \\
0 &  0
\end{bmatrix}
,~~~
T_n=
\begin{bmatrix}
0 &  u  \\
-\frac{1}{u} &  0
\end{bmatrix}
,~~~
I_n=
\begin{bmatrix}
1 &  0  \\
0 &  1
\end{bmatrix}\,.
\end{equation}
If the quadrupoles are tilted by 45$^\circ$ the 4$\times$4 transfer matrix can be written as
\begin{equation}
\label{simple_decoupling_matrix}
\overline R=R_r R_q R_r^T=\frac{1}{2}
\begin{bmatrix}
T_{n+} &  T_{n-}\\
T_{n-} &  T_{n+}
\end{bmatrix},
\end{equation}
where
\begin{equation}
R_r=\frac{1}{\sqrt 2}
\begin{bmatrix}
I_n &  I_n  \\
-I_n &  I_n
\end{bmatrix}\,,\,\,\,\,
{T_{n\pm}=T_n \pm I_n}.
\end{equation}
The beam matrix $C_{3}^{'}$ after the decoupling section is
\begin{equation}
C_{3}^{'}=\overline R C_{2}^{'} {\overline R}^T=
\begin{bmatrix}
\eta_+ \Gamma_{n+} &  \zeta \Gamma_{n-} \\
\zeta \Gamma_{n-} &  \eta_- \Gamma_{n+}
\end{bmatrix},
\label{generic_uncoupled_Mommatrix}
\end{equation}
and the 2$\times$2 sub-matrices $\Gamma_{n\pm}$ are defined through
\begin{equation}
\Gamma_{n\pm}=
\begin{bmatrix}
u &  0  \\
0 &  \pm \frac{1}{u}
\end{bmatrix},
\end{equation}
with
\begin{equation}
\label{eq7}
\eta_{\pm}=\frac{\varepsilon_n}{2}(\frac{\beta_n}{u}+\frac{u}{\beta_n} \mp 2k\beta_n)
\end{equation}
and
\begin{equation}
\label{eq8}
\zeta=\frac{\varepsilon_n}{2} (-\frac{\beta_n}{u}+\frac{u}{\beta_n})\,.
\end{equation}
Assuming that this beam matrix is diagonal, its $x$-$y$ component vanishes
\begin{equation}
\label{eq9}
\zeta \Gamma_{n-}=O_n\,
\end{equation}
solved by
\begin{equation}
\label{u_beta}
u=\pm\beta_n\,,
\end{equation}
where the positive sign indicates that $\varepsilon_x$ is made equal to $\varepsilon_1$ by decoupling and the negative sign means that $\varepsilon_y$ is made equal to $\varepsilon_1$.
We calculate the final rms emittances obtaining
\begin{equation}
\label{eq11}
\varepsilon_{x,y}=|\varepsilon_n(1 \mp k\beta_n)|\,.
\end{equation}
For a given effective solenoid fringe field strength $k_0$, the corresponding quadrupole gradients may be determined using a numerical routine, such that finally the rms emittances are equal to the eigen-emittances. If these optimized gradients are applied to remove interplane correlations produced by a different fringe strength $k_1$, the resulting rms emittances and eigen-emittances at the exit of the decoupling section are calculated as
\begin{equation}
\label{rms_emits}
\varepsilon_{x,y}=\frac{\varepsilon_n(k_1)}{2}\left|\frac{\beta_n(k_1)}{\beta_n(k_0)}+\frac{\beta_n(k_0)}{\beta_n(k_1)} \mp  2k_1 \beta_n(k_1)\right|
\end{equation}
and
\begin{equation}
\label{eq13}
\varepsilon_{1,2}=\varepsilon_n(k_1)|  1 \mp k_1 \beta_n(k_1) |
\end{equation}
with the parameter
\begin{equation}
\label{t value}
t=\frac{\varepsilon^2 \beta^2  }{\frac{\varepsilon}{\beta}(\frac{\varepsilon}{\beta}+k_0^2 \varepsilon \beta)}\frac{(k_1^2-k_0^2)^2}{4}\,.
\end{equation}
In the same way the rms Twiss parameters of a beam coupled by $k_1$ but decoupled
by $\overline{R}(k_0)$ are found from Equ.~(\ref{generic_uncoupled_Mommatrix}) as
\begin{equation}
\label{Twiss_const}
\tilde{\alpha}_x \,=\,\tilde{\alpha}_y\,=\,0,\,\,\,\,\,\,\,\,\tilde{\beta}_x\,=\,\tilde{\beta}_y\,=\,\beta_n(k_0)\,,
\end{equation}
showing that the rms Twiss parameters after decoupling do not depend on the coupling solenoid fringe strength $k_1$ if the decoupling section was set assuming a coupling strength~$k_0$.\\
We stress the very convenient feature of the generic decoupling line $\overline{R}$:
once a decoupling set of gradients has been found for the fringe field strength $k_o$, these gradients will practically decouple also beams coupled by a different strength $k_1$. This is shown in Fig.~\ref{t-graph}, which was originally presented in~\cite{Xiao_prstab}. Moreover, the Twiss parameters $\beta$ and $\alpha$ at the exit of the generic beam line $\overline{R}$ do not depend on the fringe strength as illustrated also in Fig.~\ref{ellipses}. These two features enormously facilitate the design and operation of such a round-to-flat adapter.

\section{Decoupling in the general case}
\begin{figure}[hbt]
\centering
\includegraphics*[width=80mm,clip=]{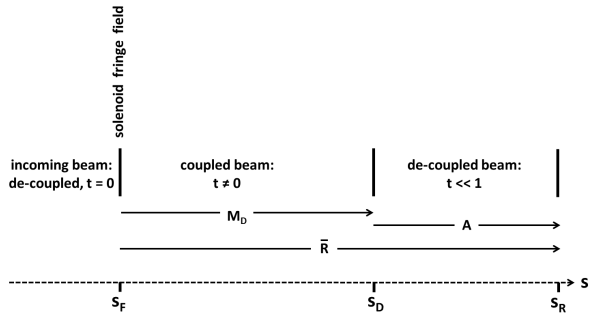}
\caption{Extension of the decoupling features of the generic beam line $\overline{R}$ to any decoupling beam line $M_D$. $S_F$ denotes the location of the initially coupling stand-alone solenoid fringe field. The arbitrary decoupling beam line $M_D$ ends at $S_D$, and the generic beam line $\overline{R}$ ends at $S_R$. The beam line $A$ does not include any x-y coupling element.}
\label{M-extension}
\end{figure}

In the previous section we derived the following ensemble $\cal{P}$ of properties of the generic decoupling line $\overline{R}$ of Equ.~\ref{simple_decoupling_matrix}:
\begin{itemize}
\item $t$ at the exit scales as $(k_1^2 - k_0^2)^2$, where $k_o$ is the assumed fringe strength and $k_1$ is the strength actually applied for the coupling. $t<<1$ holds over a wide range of $k_1$ (Equ.~\ref{t value} and Fig.~\ref{t-graph} with $B\sim k_1$).
\item the exit Twiss parameters $\beta_x$, $\alpha_x$, $\beta_y$, $\alpha_y$ do not depend on the actual fringe strength $k_1$ (Equ.~\ref{Twiss_const} and Fig.~\ref{ellipses} with $B\sim k_1$).
\item the only quantity considerably changed through the fringe strength is the transverse rms emittance partitioning $\varepsilon_x/\varepsilon_y$ (Equ.~\ref{rms_emits} and Fig.~\ref{ellipses} with $B\sim k_1$).
\end{itemize}
It must be stressed that these properties hold for both signs in Equ.~\ref{u_beta}. However, ~\cite{Xiao_prstab} found by various tracking simulations with TRACK~\cite{Peter} as well as by applying the matrix formalism, that $\cal{P}$ seems to hold for any beam line $M_D$ that provides decoupling of a beam previously coupled through a stand-alone solenoid fringe field.
This feature was not understood in~\cite{Xiao_prstab}.\\
\\
Instead it can be understood through the procedure being illustrated in Fig.~\ref{M-extension}.
Suppose there is any arbitrary beam line $M_D$ that provides decoupling. This beam line includes x-y coupling linear elements. We prolong $M_D$ by a beam line represented by the matrix
\begin{equation}
A\,=\,
\begin{bmatrix}
A_x &  O_n  \\
O_n &  A_y
\end{bmatrix}\,
\end{equation}
with the 2$\times$2 sub-matrices $A_x$ and $A_y$. $A$ must not include any x-y coupling element.\\
The resulting total beam line is the product $AM_D$. We choose for the non-coupling line $A=\overline{R}M_D^{-1}$ such that $\overline{R}=AM_D$. Care is to be taken in choosing the right sign at Equ.~\ref{u_beta} in the construction of $\overline{R}$. This is to assure that both, $M_D$ and $\overline{R}$, reduce $\varepsilon_x$ to the same of the two eigen-emittances. Choosing the wrong sign, $A$ gets an emittance exchange beam line that includes coupling elements. As shown above, at the exit of $\overline{R}$ the properties $\cal{P}$ hold. From the exit of $\overline{R}$ the Twiss parameters $\varepsilon$, $\beta$, and $\alpha$ (in both planes) are transported backwards to $S_D$ by applying $A^{-1}$ being aware that $\alpha$ and $\beta$ do not depend from the fringe strength. As $A$ does not include any x-y coupling element, neither does $A^{-1}$. Accordingly, the back-transformed Twiss parameters at $S_D$ also do not depend on the fringe strength. The same way the invariance of the Twiss parameters w.r.t. the fringe strength is kept through the back-transportation by $A^{-1}$, the weak dependence of $t(k_1)$ is back-transported \& preserved through $A^{-1}$. Since $A^{-1}$ is non-coupling, it preserves $t$.  In other words, the properties $\cal{P}$ at the exit of $\overline{R}$ are preserved during back-transportation by $A^{-1}$. As a consequence the properties $\cal{P}$ hold also at the exit of the arbitrarily chosen decoupling beam line $M_D$.
\\
These arguments are summarized in the formula
\begin{equation}
M_D\,=\,A^{-1}\overline{R}.
\end{equation}
$\overline{R}$ has the properties $\cal{P}$, which has been derived in the previous section. The matrix $A^{-1}$ does not change them since it is non-coupling. As a consequence, the properties $\cal{P}$ are also intrinsic properties of $M_D$.

\bibliography{C.Xiao}
\end{document}